\documentclass[%
 reprint,
 amsmath,amssymb,
aps,
]{revtex4-1}

\usepackage{subfigure}
\usepackage{graphicx}% Include figure files
\usepackage{dcolumn}% Align table columns on decimal point
\usepackage{bm}% bold math
\begin{document}

\preprint{APS/123-QED}

\title{Induced Waveform Transitions of Dissipative Solitons}

\author{Bogdan A. Kochetov$^1$}

\author{Vladimir R. Tuz$^{1,2}$}
\email{tvr@jlu.edu.cn; tvr@rian.kharkov.ua}
\affiliation{$^1$International Center of Future Science, State Key Laboratory on Integrated Optoelectronics, College of Electronic Science and Engineering, Jilin University, \\ 2699 Qianjin St., Changchun 130012, China}
\affiliation{$^2$Institute of Radio Astronomy of National Academy of Sciences of Ukraine, 4, Mystetstv St., Kharkiv 61002, Ukraine} 

\date{\today}

\begin{abstract}
The effect of an externally applied force upon dynamics of dissipative solitons is analyzed in the framework of the one-dimensional cubic-quintic complex Ginzburg-Landau equation supplemented by a linear potential term. The potential accounts for the external force manipulations and consists of three symmetrically arranged potential wells whose depth is considered to be variable along the longitudinal coordinate. It is found out that under an influence of such potential a transition between different soliton waveforms coexisted under the same physical conditions can be achieved. A low-dimensional phase-space analysis is applied in order to demonstrate that by only changing the potential profile, transitions between different soliton waveforms can be performed in a controllable way. In particular, it is shown that by means of a selected potential, propagating stationary dissipative soliton can be transformed into another stationary solitons as well as into periodic, quasi-periodic, and chaotic spatiotemporal dissipative structures. 
%\begin{description}
%\end{description}
\end{abstract}

\pacs{Valid PACS appear here}% PACS, the Physics and Astronomy
\maketitle

%\tableofcontents

\section{\label{intr}Introduction}
The complex Ginzburg-Landau equation (CGLE) arises in many fields of sciences including nonlinear optics, semiconductor devises, Bose-Einstein condensates, superconductivity, reaction-diffusion systems, quantum field theories and occupies a prominent place in the theory of nonlinear evolution equations \cite{Cross_RMP_1993, Aranson_RMP_2002, Rosanov_Book_2002, Malomed_Encyclopedia_2005}. Moreover, the standard cubic CGLE supplemented by higher-order nonlinear terms forms a universal mathematical model used to effectively describe diverse nonlinear phenomena possessing complex dynamical behaviors. In fact, being a nonintegrable dynamical system near a subcritical bifurcation the one-dimensional cubic-quintic CGLE admits a variety of stable localized solutions including formation of periodic and quasi-periodic patterns, as well as a spatiotemporal chaos \cite{Akhmediev_Chapter_2005}. These solutions represent different forms of dissipative solitons \cite{Akhmediev_Book1, Akhmediev_Book2, Liehr_Book}, which appear as stable localized structures existing due to a balance between gain and loss in distributed nonlinear dynamical systems far from equilibrium. Dissipative solitons may stably evolve as stationary zero-velocity solitons (so-called `plain pulse' and `composite pulse' \cite{Afanasjev_PRE_1996}) and moving solitons \cite{Fauve_PRL_1990, van_Saarloos_PD_1992, Soto-Crespo_PLA_2001}, periodically and quasi-periodically pulsating solitons with simple or more complicated behaviors \cite{Deissler_PRL_1994, Akhmediev_PRE_2001}, chaotic solitons \cite{Deissler_PRL_1994, Akhmediev_PRE_2001}, and exploding solitons, which periodically manifest explosive instabilities returning to their original waveforms after each explosion \cite{Soto-Crespo_PRL_2000, Akhmediev_PRE_2001, Soto-Crespo_PLA_2001, Cundiff_PRL_2002}. Furthermore, the cubic-quintic CGLE has multisoliton solutions \cite{Akhmediev_PRL_1997} and admits solutions in many different waveforms of fronts, sources, sinks, and bound states \cite{van_Saarloos_PRL_1990, Malomed_PRA_1990, Malomed_PRA_1991, van_Saarloos_PD_1992, Afanasjev_PRE_b, Turaev_PRE_2007}. 

The available variety of dissipative solitons can be controlled by means of an externally applied force which influences their waveforms and, thus, can change behaviors of the soliton evolution. Independently on the physical nature, the effect of an external force can be described considering soliton waveform evolution that occurs in a corresponding external potential. For instant, such a concept is used to manage solitons in nonlinear optical systems and Bose-Einstein condensates \cite{Malomed_Book}. In order to split optical spatial solitons governed by the nonlinear Schr\"odinger equation into two solitons a longitudinal defect \cite{Fratalocchi_PRE_2006}, an external delta potential \cite{Holmer_JNS_2007}, and a longitudinal potential barrier \cite{Yang_OE_2008} are applied to scatter a single soliton. Various scenarios of the dynamics of dissipative solitons interacting with a sharp potential barrier in the cubic-quintic CGLE are analyzed in \cite{He_JOSAB_2010}. Similarly, the evolution of dissipative solitons in an active bulk medium has been studied in the framework of the two-dimensional CGLE with an umbrella-shaped \cite{Yin_JOSAB_2011} and a radial-azimuthal \cite{Liu_OE_2013} potentials. 

As an example of practical systems supporting propagation of dissipative solitons, nonlinear magneto-optic waveguides can be mentioned where an externally applied magnetic field induces corresponding potential \cite{Boardman_1995, Boardman_1997, Boardman_2001, Boardman_2003, Boardman_2005, Boardman_2010}. Significant benefits of using a spatially inhomogeneous external magnetic field to adjust nonreciprocal propagation of light dissipative solitons in magneto-optic planar waveguides have been demonstrated in \cite{Boardman_Chapter_2005, Boardman_2006}. Moreover, involving an external magnetic field a new robust mechanism to perform both selective lateral shift within a group of stable dissipative solitons \cite{OptLett_2017} and their cascade replication \cite{PRE_2017} is recently proposed.

In this paper we employ the one-dimensional cubic-quintic CGLE supplemented by a linear potential term in order to reveal the effects of external force which influences upon  dissipative solitons. We demonstrate that applying particular inhomogeneous potential wells can cause nontrivial transitions between different soliton waveforms. These transitions appear when the waveforms coexist within the same parameter space of the CGLE \cite{Afanasjev_PRE_1996, Akhmediev_PRE_2001, Soto-Crespo_PLA_2001}. The goal of the paper is to perform a formalized description of peculiarities of the induced waveform transitions of dissipative solitons and related phenomena.

The rest of the paper is organized as follows: In Sec.~\ref{mod} we formulate a common mathematical model of dissipative solitons, where a linear potential term is added to the CGLE to perform some manipulations upon the solitons. The existence of regular waveform transitions of dissipative solitons caused by the \textit{weak} potential is demonstrated in Sec.~\ref{rwt}. In Sec.~\ref{iwt} we present and discuss peculiarities of irregular waveform transitions of dissipative solitons and the related effect that appear under an influence of the \textit{strong} potential. Conclusions and final remarks summarize the paper in Sec.~\ref{concl}.

\section{\label{mod} Model of controllable dissipative solitons}
We consider $1+1$ dimensional cubic-quintic CGLE supplemented by a linear potential term written in the form
\begin{multline}
\label{CQCGLE}
\mathrm{i}\frac{\partial\Psi}{\partial z}+\mathrm{i}\delta\Psi+\left(\frac{D}{2}-\mathrm{i}\beta\right)\frac{\partial^2\Psi}{\partial x^2}+\left(1-\mathrm{i}\varepsilon\right)\left|\Psi\right|^2\Psi \\ - \left(\nu-\mathrm{i}\mu\right)\left|\Psi\right|^4\Psi + Q(x,z)\Psi = 0,
\end{multline}
where $\Psi\left(x,z\right)$ is a complex amplitude of transverse $x$ and longitudinal $z$ coordinates, $D$ is the group velocity dispersion coefficient, $\delta$ is a linear absorption, $\beta$ is a linear diffusion, $\varepsilon$ is a nonlinear cubic gain, $\nu$ accounts for the self-defocusing effect due to the negative sign, and $\mu$ defines quintic nonlinear losses. The linear potential $Q(x,z)$ is considered to be in the form of arranged potential wells whose depth is varied along the longitudinal coordinate $z$. We express the profile of each potential well through the hyperbolic tangent function and write down the whole potential as some superposition of the wells
\begin{equation}
\label{Q} Q(x,z) = \sum_{i=1}^{N} M_i\tanh\left(\frac{q_i(z)}{\sqrt{(x-x_i)^2+d_i^2}}\right),
\end{equation}
where $N$ is the number of wells, $x_i$ is the transverse coordinate of the $i$-th potential well peak or deep, the profile function $q_i(z)$ defines the variation of the $i$-th potential well along the longitudinal coordinate $z$, and $M_i$ and $d_i$ are some constants which should be selected based on a particular physical problem under consideration.

We assume that the complex amplitude $\Psi(x,z)$ satisfies to the periodic boundary condition
\begin{equation}
\label{PBC}
\Psi(x,z) = \Psi(x+L_x,z),~~~~(x,z)\in\mathbb{R}\times\left[0,+\infty \right),
\end{equation}
and the initial condition 
\begin{equation}
\label{IC}
\Psi(x,0) = \Psi_0(x),~~~~x\in\mathbb{R},
\end{equation}
where $L_x$ is some given number and the initial amplitude $\Psi_0(x)$ satisfies to the periodic condition \eqref{PBC} as well.

The system \eqref{CQCGLE} is nonconservative since its solutions (dissipative solitons) depend on an energy supplied to the system. Indeed, the following continuity equation can be derived using Eq.~\eqref{CQCGLE}
\begin{equation}
\label{CE}
\dfrac{\partial\rho}{\partial z}+\dfrac{\partial j}{\partial x}=p,
\end{equation}
where $\rho=\left|\Psi(x,z)\right|^2$ is the energy density. The corresponding flux $j$ and the density of energy generation $p$ are defined as follows
\begin{equation}
\label{j}
j=\dfrac{\mathrm{i}D}{2}\left(\Psi\Psi_x^*-\Psi_x\Psi^*\right),
\end{equation}
\begin{multline}
\label{p}
p=\beta\left(\left|\Psi\right|^2_{xx}-2\left|\Psi_x\right|^2 \right) \\ -2\left(\delta\left|\Psi\right|^2-\varepsilon\left|\Psi\right|^4+\mu\left|\Psi\right|^6\right).
\end{multline}

Having integrated the energy density $\rho$ and the density of energy generation $p$ over the transverse coordinate $x$ we get two soliton parameters (moments) to observe their evolution along the $z$-axis
\begin{equation}
\label{mom}
E(z) = \int_{-\infty}^{\infty} \left|\Psi(x,z)\right|^2 dx,~~P(z) = \int_{-\infty}^{\infty} p(x,z)\,dx.
\end{equation}
These two parameters represent the soliton energy $E(z)$ and total generated energy $P(z)$ as functions of the propagation distance $z$, respectively.

\section{\label{na}Numerical analysis}
Further we solve the problem \eqref{CQCGLE}-\eqref{IC} numerically using the pseudospectral approach and exponential time differencing method of second order \cite{Beylkin_JCP_1998, Cox_2002}. Since we seek solutions $\Psi(x,z)$ to Eq.~\eqref{CQCGLE} in the form of dissipative solitons which are structures localized in space, we assume that the computational domain can be reduced to the finite rectangular occupying the area $\left[-L_x/2,L_x/2\right]\times\left[0,L_z\right]$. In our numerical calculations, we sample the computational domain along the $x$-axis with $2^{10}$ discretization points to compute the fast Fourier transform with respect to the $x$ coordinate. The distance along the $z$-axis we sample using the step $\Delta z=10^{-3}$. The length of the simulation area along the $z$-axis is chosen so as to ensure the completion of all intermediate unstable stages appearing between distinct waveforms transitions. 

The computational scheme in the Fourier domain for updating the unknown complex amplitude along the longitudinal coordinate $z$ is written in the form
\begin{multline}
\label{shc}
\hat{\Psi}_{n+1}=\hat{\Psi}_{n}e^{\sigma\Delta z}+\hat{\mathcal{N}}_n\dfrac{\left(1+\sigma\Delta z\right)e^{\sigma\Delta z}-1-2\sigma\Delta z}{\sigma^2\Delta z}\\
+\hat{\mathcal{N}}_{n-1}\dfrac{1+\sigma\Delta z-e^{\sigma\Delta z}}{\sigma^2\Delta z},
\end{multline}
where $\sigma=-\delta-\left(\beta+\mathrm{i}D/2\right)k^2$ is a spectral parameter, $\Psi_n=\Psi(x,z_n)$, $\mathcal{N}_n=\mathcal{N}\left(\Psi_n,x,z_n\right)$, $z_n=n\Delta z$, and $\mathcal{N}$ stands for the nonlinear part of Eq.~\eqref{CQCGLE}
$$\mathcal{N}(\Psi,x,z)=\left(\left(\mathrm{i}+\varepsilon\right)\left|\Psi\right|^2\right.\left.-\left(\nu-\mathrm{i}\mu\right)\left|\Psi\right|^4+Q(x,z)\right)\Psi.$$
The circumflex denotes the Fourier transform with respect to the coordinate $x$, i.e. $\hat{\Psi}(k,z)= \mathcal{F}\left\lbrace\Psi(x,z)\right\rbrace$, $\hat{\mathcal{N}}= \mathcal{F}\left\lbrace \mathcal{N}(\Psi,x,z)\right\rbrace$.

In order to excite a stable dissipative soliton a wide variety of functions whose profiles are close to the existing soliton waveform can be used as the initial condition \eqref{IC}. In particular, in all our numerical calculations the initial plane pulse soliton is evolved from the simple waveform function $\mathrm{sech}(x)$.

Hereinafter we consider the case of anomalous group velocity dispersion, i.e. $D=1$. Moreover, in the potential \eqref{Q} we put $M_i=2$ and $d_i=1$, $i=1,\ldots,N$, while its dependence on the longitudinal coordinate $z$ is set to be in the form of a simple piecewise constant function
\begin{equation}
\label{q}
q_i(z)=A_i h(z-a_i)-B_i h(z-b_i),
\end{equation}
where $h(\cdot)$ is the Heaviside step function and $A_i$, $a_i$, $B_i$, and $b_i$ are some real numbers.

\subsection{\label{rwt}Regular waveform transitions}
\begin{figure}[htbp]
\centering
\includegraphics[width=\linewidth]{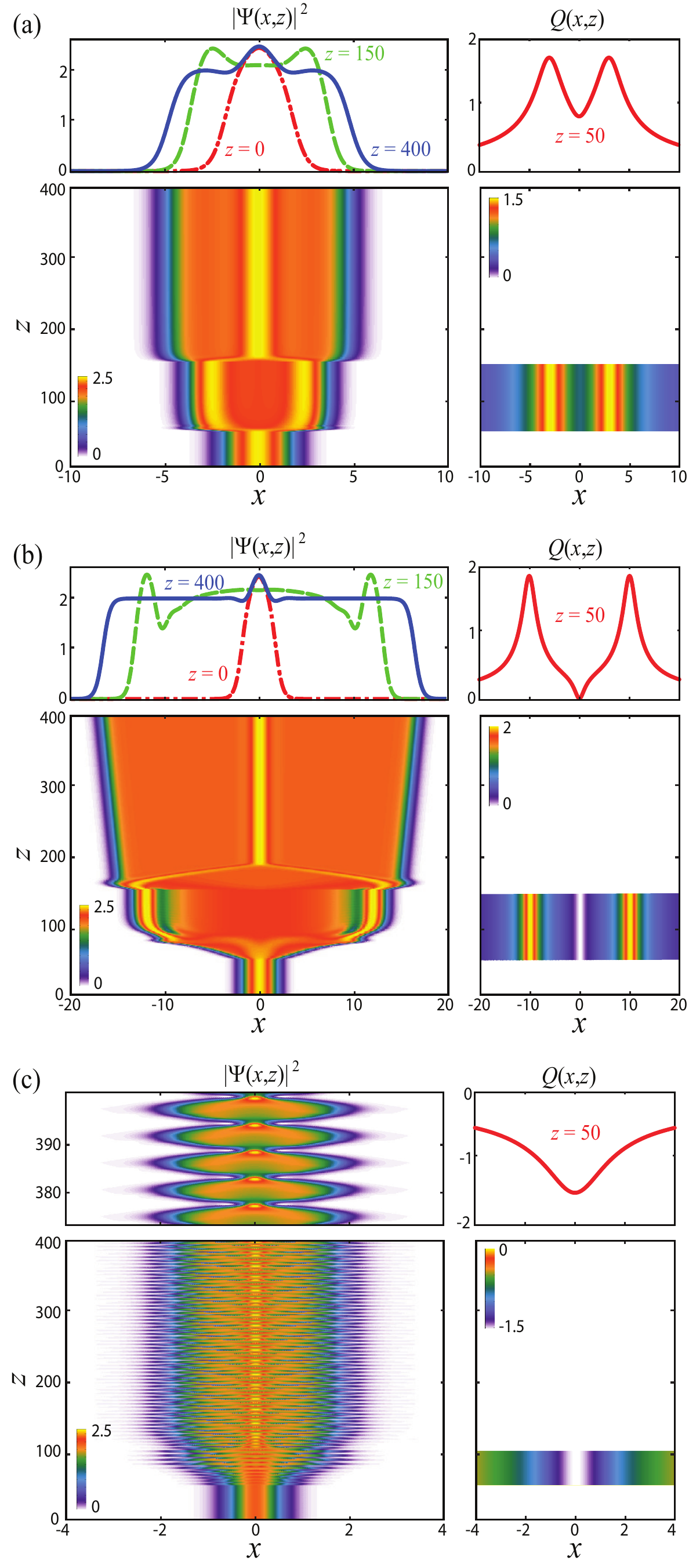}
\caption{Illustration of regular waveform transitions of plain pulse to (a)~composite pulse, (b)~composite pulse with moving fronts, and (c)~pulsating soliton that appear under an influence of different potentials $Q$. Intensity distribution of squared absolute value of complex amplitude and potentials are presented in left and right panels, respectively. Parameters are: (a) $\beta=0.5$, $\delta=0.5$, $\mu=1$, $\nu=0.1$, $\epsilon=2.53$, $A_i=B_i\in\{1,-0.2,1\}$, $a_i=50$, $b_i=150$, $N=3$, $x_i\in\{-3,0,3\}$; (b) $\beta=0.5$, $\delta=0.5$, $\mu=1$, $\nu=0.1$, $\epsilon=2.53$, $A_i=B_i\in\{1,-0.2,1\}$, $a_i=50$, $b_i=150$, $N=3$, $x_i\in\{-10,0,10\}$; (c) $\beta=0.08$, $\delta=0.1$, $\mu=0.1$, $\nu=0.07$, $\epsilon=0.75$, $A_1=B_1=-1$, $a_1=50$, $b_1=100$, $N=1$, $x_1=0$.}
\label{fig_1}
\end{figure}

In this section we consider an evolution of dissipative solitons being under an influence of the locally applied weak potential in order to reveal peculiarities of the soliton transitions between different waveforms implying these soliton waveforms coexist in the same equation parameter space. In particular, parameters of Eq.~\eqref{CQCGLE} are chosen so as to admit waveform coexistence of plain pulse, composite pulse and pulsating solitons. A set of possible waveform transitions influenced by the weak potential is demonstrated in details in Fig.~\ref{fig_1}. In each panel of this figure the cross-section profiles (top) and the intensity plots (bottom) of the squared absolute value of complex amplitude $|\Psi(x,z)|^2$ (left) and the potential $Q(x,z)$ (right) are presented. 

First, the stages of soliton transition from a plain pulse to a composite pulse are presented in Fig.~\ref{fig_1}(a). At the section $z=0$ the plain pulse comes into existence and then propagates freely through the zero potential up to the section $z=50$, where the two-humped potential $Q$ abruptly arises. This potential influences upon the plain pulse soliton transiting its waveform to another stationary state allowable under such an applied potential. Next, at the section $z=150$, the potential $Q$ becomes to be zero, and the waveform perturbed by the potential transits to the composite pulse, which coexists with the initial plain pulse under the same equation parameters.   

From the viewpoint of the theory of dynamical systems, both plain pulse and composite pulse are $z$-independent or stationary solutions, which are associated with two stable isolated fixed points existing in the infinite-dimensional phase space of the system \eqref{CQCGLE}. For each such fixed point, certain set of initial conditions forms a basin of attraction. Thus, the applied potential influences strongly enough upon the solition to transit the soliton waveform from the vicinity of the plain pulse fixed point to the basin of attraction of the composite pulse fixed point.

Therefore, taking this property into account one can controllably perform desired waveform transition of the dissipative solitons between all basins of attraction which coexist in the same equation parameter space. This ability is further demonstrated in Fig.~\ref{fig_1}(b) where the transition from the vicinity of the plain pulse fixed point to the basin of attractor of the composite pulse with moving fronts is realized. This transition is caused by the two-humped potential $Q$, whose profile is just slightly different from the previously discussed one. 

A new set of the equation parameters gives us another example of the coexistence of plain pulse and pulsating dissipative solitons. In Fig.~\ref{fig_1}(c) we demonstrate waveform transition of the stationary plain pulse to the pulsating soliton that periodically changes its waveform along the longitudinal coordinate $z$. This transition is induced by the repulsive (defocusing) potential $Q$. In this figure five steady-state pulsations are zoomed in and presented in the upper left corner of the panel, where one can notice that the period of pulsations is approximately equal to $5$ on the $z$-axis scale. 

\begin{figure}[htbp]
\centering
\includegraphics[width=\linewidth]{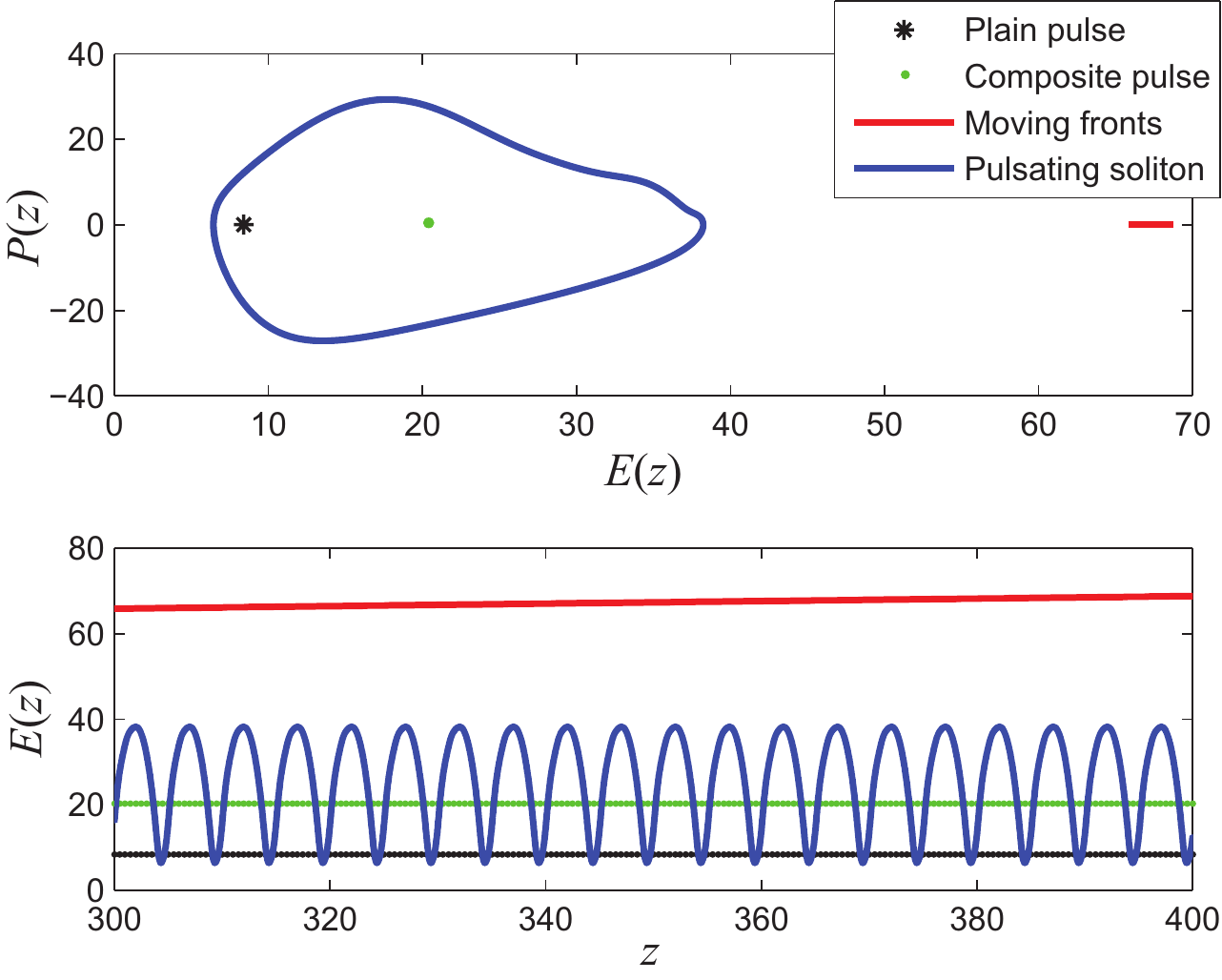}
\caption{Phase trajectories in spaces $\mathcal{P}_1$ (top) and $\mathcal{P}_2$ (bottom) describing soliton behaviors related to the plain pulse, composite pulse, composite pulse with moving fronts, and pulsating soliton.}
\label{fig_2}
\end{figure}

Such a pulsating soliton can be considered as a limit cycle in a phase space of the dynamical system \eqref{CQCGLE}. In order to visualize this limit cycle and other attractors of the system as well as to identify soliton behaviors along the $z$-axis we perform the low-dimensional phase-space analysis which is based on the flow projections onto a pair of two-dimensional spaces. They are defined as follows $\mathcal{P}_1=\{\left(E(z),P(z)\right),\in\mathbb{R}^2\}$ and $\mathcal{P}_2=\{\left(z,E(z)\right),\in\mathbb{R}^2\}$, where components of the projections are the soliton energy parameters \eqref{mom} and the longitudinal coordinate $z$. Trajectories in the two-dimensional phase spaces that correspond to the mentioned above attractors are plotted in Fig.~\ref{fig_2}. Both plain pulse and composite pulse are represented by a single point with zero total generated energy and horizontal line in the spaces $\mathcal{P}_1$ and $\mathcal{P}_2$, respectively. For the composite pulse with moving fronts, the trajectory in the space $\mathcal{P}_1$ degenerates into a straight line while the corresponding straight line in the space $\mathcal{P}_2$ has positive inclination. Finally, for the pulsating soliton, points in the space $\mathcal{P}_1$ always lies on a closed loop (cycle) while the soliton energy along the $z$-axis in the space $\mathcal{P}_2$ appears as a periodic function.

In Fig.~\ref{fig_1} we have demonstrated three particular examples of induced waveform transitions of dissipative solitons, where the initial plain pulse waveform transits to another ones. In fact, the process of soliton waveform transition is invertible. Indeed, in order to invert the considered direct waveform transitions one can apply to the composite pulse or pulsating soliton a quite high single-well potential (is not presented here, for illustration of the invert transitions see Ref.~\cite{Metanano_2017}). Moreover, if for a fixed set of equation parameters, the dynamical system \eqref{CQCGLE} has two or more attractors then waveform transitions between each pair of the basin of attraction can be induced by applying a suitable linear potential $Q$ with a finite support. In other words, any induced waveform transitions between coexisted stable waveforms are possible.

The considered cases demonstrate a variety of the induced waveform transitions to different waveforms. However, all these transitions are performed according to the same scenario. An initially excited soliton approaches its attractor until an appropriate external potential is applied. The potential inevitably influences upon the soliton waveform and changes it drastically. Then waveform transition is performed in two stages. The first one is a relatively short transient period which begins as soon as the potential is applied and it  finishes when the perturbed waveform reaches a new steady state. During the second stage a new steady state of the perturbed waveform exists. It continues as long as the potential is applied. When the potential is switched back to zero the perturbed waveform returns from the induced steady state to some attractor of the system \eqref{CQCGLE} with zero potential. For many sets of parameters, the system \eqref{CQCGLE} has more than one attractor. Therefore, depending on which basin of attraction contains the perturbed waveform it approaches the initial attractor or another one. In all considered cases we chose the potential in such a way that the perturbed steady state waveform does not belong to the initial basin of attraction. Thus, switching off the potential unavoidably changes the initial soliton waveform to another one. Remarkably, this mechanism of waveform changing is stable to small variations of the potential profile, and waveform transitions have the same outcome for similar potential profiles. Therefore, we distinguish these waveform transitions as \textit{regular} ones. 

\subsection{\label{iwt}Irregular waveform transitions}
\begin{figure}[htbp]
\centering
\includegraphics[width=\linewidth]{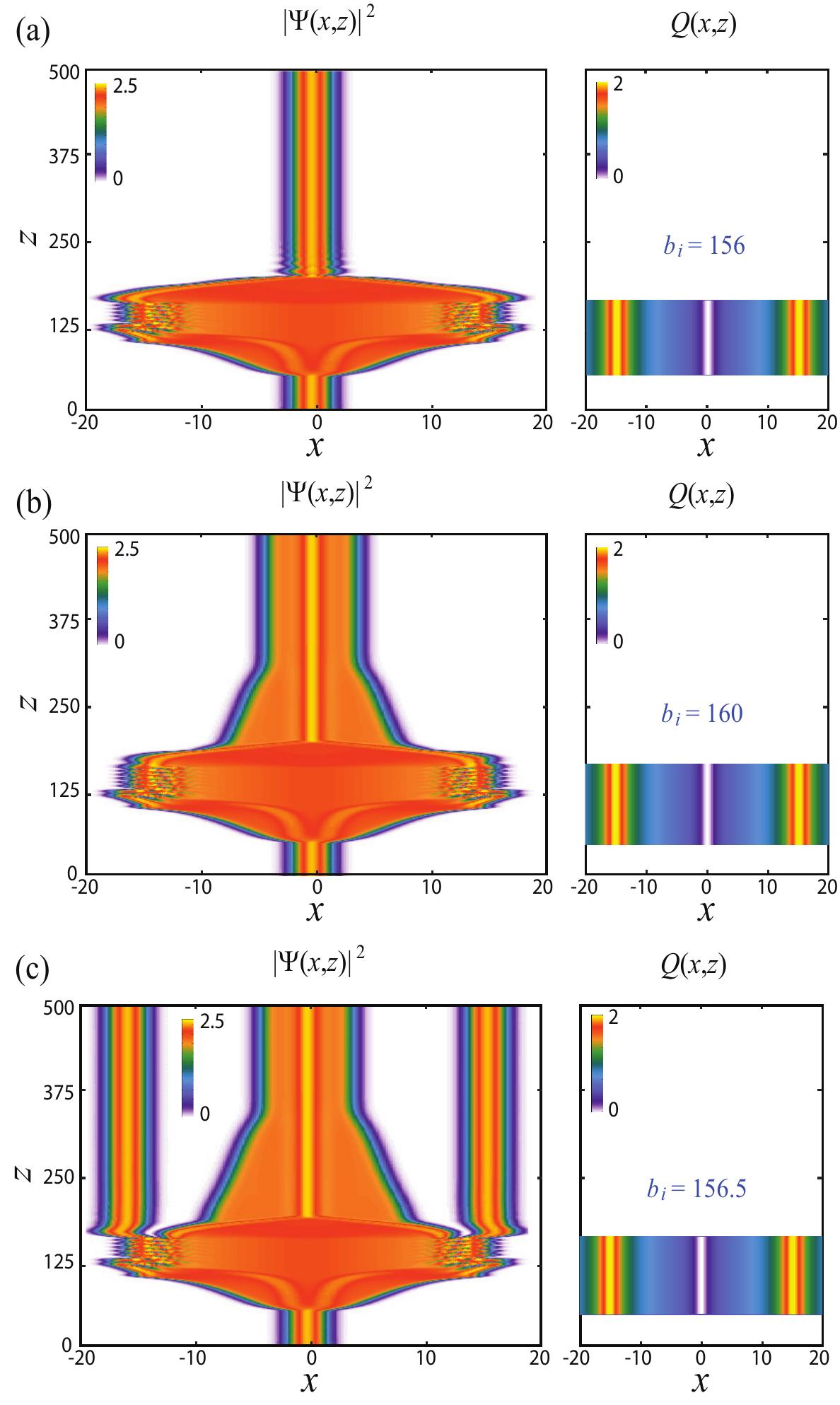}
\caption{Illustration of drastically different outcomes of the irregular waveform transitions of dissipative soliton to (a) plain pulse, (b) composite pulse, and (c) three noninteracting solitons that appear under an influence of the potential $Q$, which is switched off at different points along the $z$ axis. Intensity distribution of squared absolute value of complex amplitude and potentials are presented in left and right panels, respectively. Parameters of the equation are: $\beta=0.5$, $\delta=0.5$, $\mu=1$, $\nu=0.1$, and $\epsilon=2.52$; parameters of the potential are: $N=3$, $x_i\in\{-15,0,15\}$, $B_i=A_i\in\{1.2,-0.4,1.2\}$, $a_i=50$.}
\label{fig_3}
\end{figure}

In previous Section we have demonstrated an effect of influence of the relatively weak potential resulting in a set of particular regular waveform transitions of the dissipative solitons. The potential `weakness' means that a perturbed soliton waveform reaches its new stationary steady state after performing some finite transient stage before stabilization. However, it is revealed that applying a stronger potential can result in some nontrivial waveform transitions when a perturbed soliton waveform remains to be unstable without achieving any stationary profile, i.e. it becomes to be a pulsating soliton. The transitions that occur through pulsating waveforms and whose outcome is very sensitive to small variations in the potential profile are further considered as \textit{irregular} waveform transitions. They are presented in Fig.~\ref{fig_3}, where one can see three possible outcomes of the irregular waveform transitions from the same initial plain pulse. Remarkably, these three transitions appear under the influence of potentials which differ in the value of only one parameter $b_i$ of the function \eqref{q}. In fact, this parameter defines the longitudinal coordinate $z$ at which the potential is switched off.  

All considered irregular waveform transitions are performed according to the same scenario. A plain pulse comes into existence and freely propagates until the potential is switched on at the coordinate $z=50$. This potential influences upon the soliton changing its waveform from a plain pulse to a periodically pulsating soliton. When a particular potential is switched off the periodic waveform pulsations vanish and the soliton acquires a stationary waveform. Remarkably, after the potential removal the soliton has an alternative to acquire the form between different profiles from a set of waveforms coexisted in the same equation parameter space. In particular, for the chosen equation parameters, the pulsating soliton transits either to a single plain pulse (Fig.~\ref{fig_3}(a)), single composite pulse (Fig.~\ref{fig_3}(b)), or two plain and one composite pulses (Fig.~\ref{fig_3}(c)). In fact, the releasing from pulsations depends drastically on the potential parameters and phase of periodical pulsations at which the potential is switched off.

\begin{figure}[htbp]
\centering
\includegraphics[width=\linewidth]{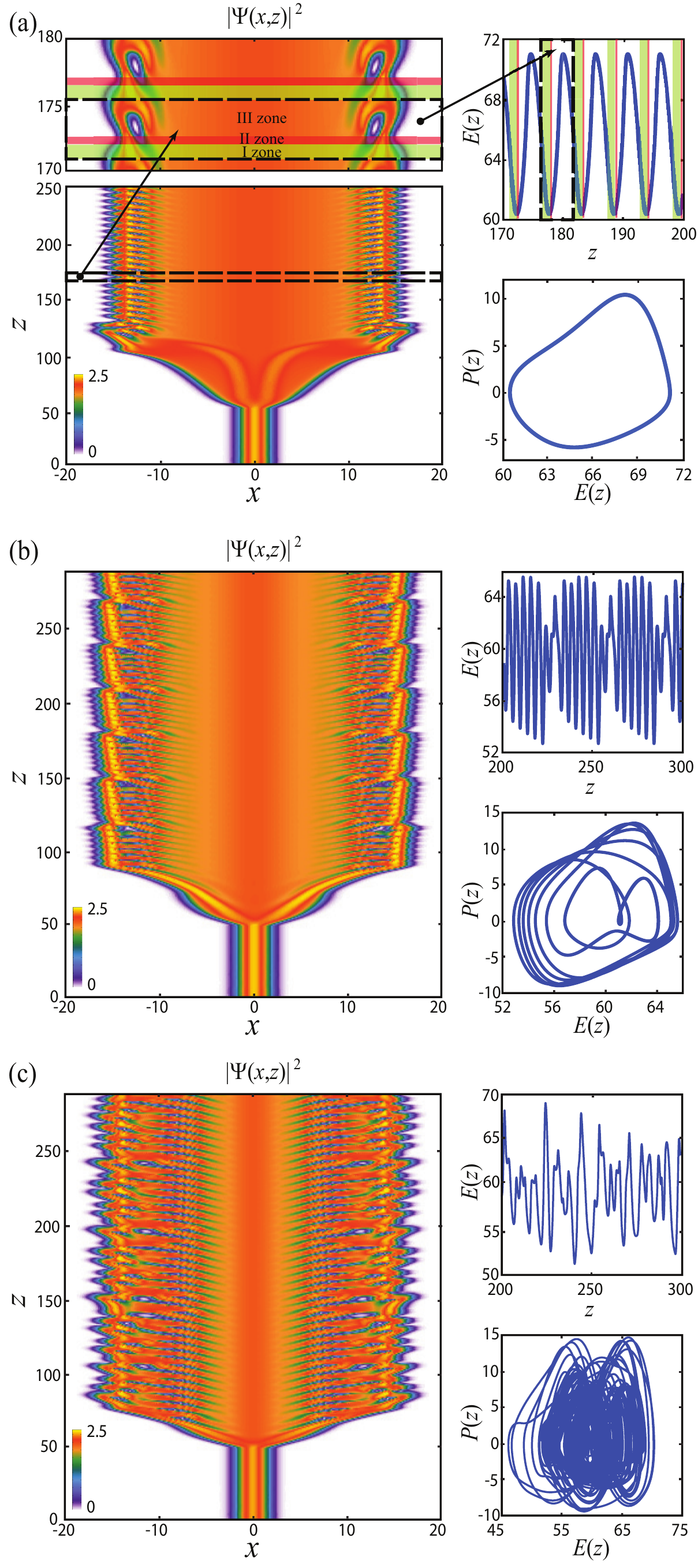}
\caption{Intensity distribution of squared absolute value of complex amplitude (left panels) and phase spaces $\mathcal{P}_1$ and $\mathcal{P}_2$ (right panels). Waveform evolution plain pulse to (a)~periodically, (b)~quasi-periodically, and (c)~chaotically pulsating solitons which coexist under the same equation parameters. Parameters of the equation are: $\beta=0.5$, $\delta=0.5$, $\mu=1$, $\nu=0.1$, and $\epsilon=2.52$; parameters of the potential are: $N=3$, $x_i\in\{-15,0,15\}$, $B_i\in\{0,0,0\}$, $a_i=50$, and (a)~$A_i\in\{1.2,-0.4,1.2\}$; (b)~$A_i\in\{1.2,-0.8,1.2\}$; (c)~$A_i\in\{1.2,-1.8,1.2\}$.}
\label{fig_4}
\end{figure}

In order to explain these multiple outcomes of irregular waveform transitions we discuss peculiarities of the soliton pulsations which are presented in Fig.~\ref{fig_4}. In each panel of this figure the intensity plot of the squared absolute value of complex amplitude $|\Psi(x,z)|^2$ (left) and flow projections onto two-dimensional spaces $\mathcal{P}_1$ and $\mathcal{P}_2$ (right) are presented. The soliton waveform evolution from a plane pulse to a periodically pulsating soliton is presented in Fig.~\ref{fig_4}(a). The unstabilized waveform appears as soon as the potential is imposed and pulsations continue to exist while the potential is in action. The pulsations demonstrate a perfect periodic behavior which arises along the $z$-axis with the period being approximately 5. One period of pulsations is outlined in Fig.~\ref{fig_4}(a) by black dashed lines. The periodic pulsations are also confirmed by the flow projections onto the spaces $\mathcal{P}_1$ and $\mathcal{P}_2$. One can see that soliton energy $E$ possesses an exactly periodic behavior along the longitudinal coordinate $z$ and the trajectory in the space $\mathcal{P}_1$ appears as a cycle which repeats itself indefinitely as long as the potential is applied.

In each period of the soliton pulsations we further distinguish three non-overlapping zones where there are corresponding instantaneous waveform profiles from which a transition to three mentioned stationary waveforms occurs. These zones are denoted by Roman numerals I, II, and III in the upper fragment of Fig.~\ref{fig_4}(a). Each such distinguished zone within the period contains all the soliton waveforms belonging to the same basin of attraction. In particular, if the potential is switched off at the phase of periodical pulsations being within zones I, II, and III, all instantaneous waveforms transit to steady state forms belonging to the basin of attraction of the plain pulse, three noninteracting solitons, and composite pulse, respectively. 

Pulsating solitons can demonstrate more complicated behaviors when the applied potential becomes stronger. In particular, it is found out that by choosing an appropriate potential, period-1 pulsations can be changed to quasi-periodical ones. This outcome occurs when a soliton waveform perturbed by a strong enough potential does not reach any stabilized form being neither stationary not simply pulsating soliton. Such an example of the waveform evolution of a plain pulse to a quasi-periodically (with period-8) pulsating soliton is presented in Fig.~\ref{fig_4}(b). The applied potential influences upon the soliton in such a way that both original waveform and energy parameters recur after every eighth pulsation, while the trajectory in the space $\mathcal{P}_1$ traces an eight-loops cycle. This cycle repeats itself indefinitely as long as the potential is applied.

Further increasing the height of potential wells results in a pulsating soliton becomes to be a chaotic one. A numerical example of such chaotic pulsations is demonstrated in Fig.~\ref{fig_4}(c). The final soliton waveform obtained from a plain pulse continuously evolves along the $z$-axis and never repeats itself remaining to be smooth and localized, while the dependence of the soliton energy $E$ on the coordinate $z$ possesses an oscillating behavior without any obvious repetitions. The trajectory in the space $\mathcal{P}_1$ densely fills a finite region manifesting behaviors of a strange attractor. 

Both quasi-periodic and chaotic pulsating solitons exist while the potential is applied. As soon as the potential is abruptly switched off, the soliton evolves back to a particular unperturbed stationary waveform. Similarly to the above discussed case of the periodically pulsating soliton, the perturbed quasi-periodically and chaotically pulsating solitons can acquire different profiles from coexisted ones in the same equation parameter space. 

\begin{figure}[htbp]
\centering
\includegraphics[width=\linewidth]{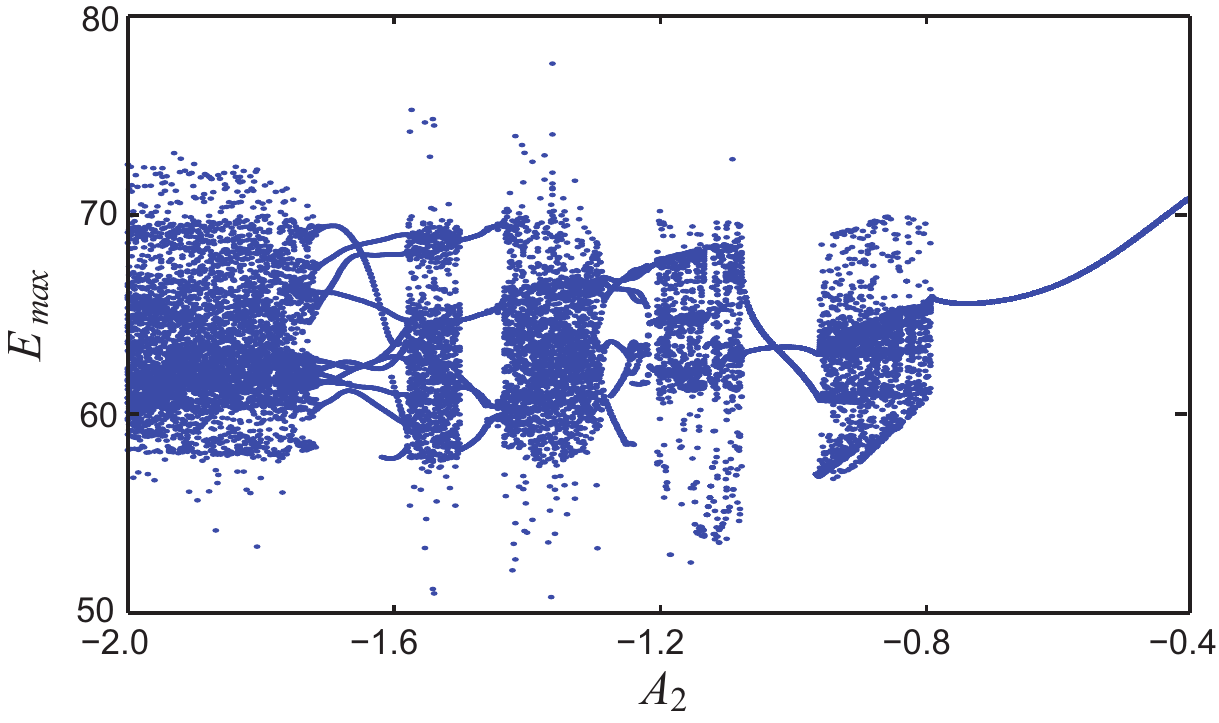}
\caption{Local maxima of pulsating soliton energy versus height of central potential well. Parameters of the equation are: $\beta=0.5$, $\delta=0.5$, $\mu=1$, $\nu=0.1$, and $\epsilon=2.52$; parameters of the potential are: $N=3$, $x_i\in\{-15,0,15\}$, $B_i=0$, $a_i=50$, $A_1=A_3=1.2$.}
\label{fig_5}
\end{figure}

Additionally we have calculated all local maxima of the pulsating soliton energy as a function of the height of applied potential (Fig.~\ref{fig_5}). This dependence can be interpreted as the one-dimensional Poincar\'e map, where a particular solution parameter (maxima of the soliton energy) is a function of an equation parameter (potential height). We calculate the map by varying the height of the central potential well $A_2$ within the interval $[-2,-0.3]$ having excited the stationary plain pulse as in the previous irregular cases. For each fixed $A_2$ we track the soliton propagation until any transients related to the switching on the potential have decayed and the soliton evolves to its perturbed waveform. Then we calculate the soliton energy $E$ as a function of the longitudinal coordinate $z$ and find all its local maxima $E_{max}$. For period-$s$ pulsating solitons we find $s$ separate points and plot them in the graph for each value of the central well height $A_2$. A chaotic soliton generates infinite sequence of different numbers distributed within some finite interval that corresponds to a continuous vertical line in the graph. In the final graph there are five domains where chaotic solitons exist. In the remaining four domains an appearance of quasi-periodic solitons is admitted, while period-$1$ pulsating soliton can exist under the condition $A_2>-0.79$. All domains possess extremely sharp boundaries indicating that both direct and inverse transitions to  chaotic waveforms happen abruptly when the potential parameter $A_2$ varies continuously.

\section{\label{concl}Conclusions}
We have considered waveform transitions of dissipative solitons induced by application of a spatially inhomogeneous potential to the system. The waveform transitions can occur only in the system having at least two attractors for the chosen equation parameters. It is demonstrated that an  appropriate potential influences upon the soliton changing an initially excited soliton waveform to a variety of perturbed waveforms which belong to different basins of attraction. Switching off the potential causes an evolution of the perturbed waveforms to another attractors. We have distinguished two types of induced waveform transitions, namely regular and irregular ones. If the outcomes of induced waveform transitions are stable with respect to small variations of the applied potential and do not depend on a particular point at witch the potential is switched off then the waveform transitions are considered to be regular. Otherwise they are irregular manifesting periodic, quasi-periodic, and chaotic behaviors. Both types of the induced waveform transitions are invertible. However, different outcomes of the direct irregular waveform transitions can be inverted by applying the same potential function. Discussed mechanism can found an application in the practical systems supporting dissipative solitons to manage their propagation.

\bibliography{waveform_transitions}

\end{document}